\documentclass{elsart}

\usepackage{graphicx,natbib,amssymb,lineno,hyperref}

\newcommand{\arcs}{$^{\prime\prime}$}

\newcommand\apj{{ApJ }}%
\newcommand\aj{{AJ }}%

\begin{document}

\begin{frontmatter}
\title{The first light-curve analysis of eclipsing binaries observed by the INTEGRAL/OMC}

\author{P. Zasche}
\ead{zasche@sirrah.troja.mff.cuni.cz}
\address{Astronomical Institute, Faculty of Mathematics and Physics,
 Charles University Prague, CZ-180 00 Praha 8, V Hole\v{s}ovi\v{c}k\'ach 2, Czech Republic}

\begin{abstract}
Three Algol-type binaries in Cygnus constellation were selected for an analysis from a huge database of
observations made by the INTEGRAL/OMC camera. These data were processed and analyzed, resulting in a first
light-curve study of these neglected eclipsing binaries. The temperatures of the primary components range from
9500~K to 10500~K and the inclinations are circa 73$^\circ$ (for PV~Cyg and V1011~Cyg), while almost 90$^\circ$
for V822~Cyg. All of them seem to be main-sequence stars, well within their critical Roche lobes. Nevertheless,
further detailed analyses are still needed.
\end{abstract}


\end{frontmatter}

\section{Introduction}

The INTEGRAL (INTErnational Gamma-Ray Astrophysics Laboratory) satellite was launched on 17 October 2002. Since
then, there were many observations of the gamma-ray and X-ray sources obtained. Thanks to the OMC (Optical
Monitoring Camera), there was collected also a large database of the observations in the visual passband.

For a detailed description of the camera, its efficiency and parameters, see e.g. \cite{2004ESASP}. Due to its
relatively large field of view (almost 5 $\times$ 5 degrees) there were observed also many photometric variable
stars near the gamma and X-ray sources as a by-product. The main advantage of the OMC data is the duration of
the continuous time series of observations, which could reach up to a few days without any interruption. This
could not be achieved by the ground-based telescopes. Due to this fact, there could be discovered also very slow
variables, or the minima of very slow Algol-type eclipsing binaries (hereafter EBs) could be catched during one
observation run.

Despite the fact that the older OMC observations are available on the internet\footnote{see
$\mathrm{https://sdc.laeff.inta.es/omc}$}, the data mining and the analysis are still very rare. Regarding the
EBs, there were only one paper about the system V435 Cas (see \citealt{V435Cas2007}) and the collection of 236
minima timings of EBs by \cite{Sobotka2007}. The analyses of the light curves of eclipsing binaries have not
been published so far.

\section{Analysis of the individual systems}

All observations of the selected systems were carried out by the same instrument (50mm OMC telescope) and the
same filter (Johnson's V filter). Time span of the observations ranges from November 2002 to July 2006. A
transformation of the time scale has been done following the equation $Julian Date - ISDC Julian Date =
2,451,544.5$. Only a few outliers from each data set were excluded. The {\sc Phoebe} programme (see e.g.
\citealt{Prsa2005}), based on the Wilson-Devinney algorithm \citep{Wilson1971}, was used.

Due to the missing information about the stars, and having only the light curves in one filter, many of the
parameters have to be fixed. At first, the temperature of the secondary component was fixed (according to the
estimated spectral type). The "Detached binary" mode was used for computing and the eccentricity was set to 0
(circular orbit). The limb-darkening coefficients were interpolated from van~Hamme's tables (see
\citealt{vanHamme1993}). The values of gravity brightening and bolometric albedo coefficients were set at their
suggested values for convective atmospheres (see \citealt{Lucy1968}), i.e. $G_1 = G_2 = 0.32$, $A_1 = A_2 =
0.5$. No third light was assumed: $l_3 = 0$.

\subsection{PV~Cyg}

The first system is PV~Cyg (= AN~93.1928 = GSC 03137-03117,
$R.A.=19\mathrm{^h}$~56$\mathrm{^m}$~29$\mathrm{^s}$, $Decl.=+37\mathrm{^\circ}$~43$^\prime$~08\arcs, J2000.0,
$B_{max} = 12.7$~mag). This star was discovered to be a variable by \cite{PVCyg1928} and its designation as
PV~Cyg was introduced by \cite{PVCyg1933}. Since then, there was no detailed analysis of the system performed,
only the times of minima observations were published. The only rough estimation of its spectral types as
A1+[G6IV] were published by \cite{Svechnikov1990}, but it is not very reliable.

\begin{figure}
 \centering
 \includegraphics[width=13.8cm]{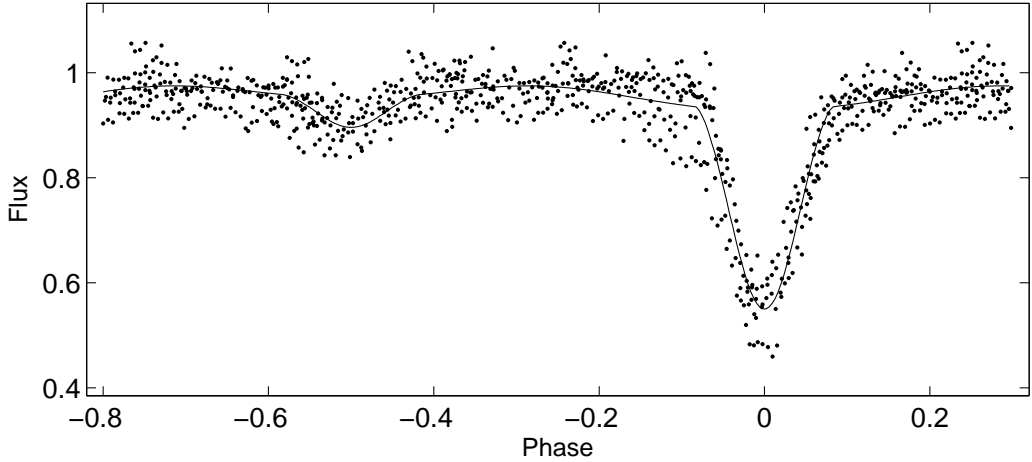}
 \caption{The $V$ light curve of PV~Cyg.}
 \label{FigPVCyg}
\end{figure}

According to these spectral types, the temperature of the primary should be circa 9100~K and the temperature of
the secondary about 5700~K (according to \citealt{Hec1988}). Due to this fact, the temperature of the secondary
was fixed $T_2 = 5700$~K and $T_1$ was calculated (from its starting value 9100~K). The final fit is plotted in
Fig.\ref{FigPVCyg}, while all the relevant parameters of the fit are presented in Table \ref{TableBig}. In this
table the ephemeris $JD_0$ and $P$ are written together with the inclination $i$, the mass ratio $q$, the radii
ratio $r_1 / r_2$, the temperature ratio $T_1/T_2$, the luminosity ratio $L_1/L_2$ and the Kopal's potentials
$\Omega_i$, respectively.

Altogether there were obtained 795 observations of the star. As one can see from Table \ref{TableBig}, the
radius of the secondary is larger than the radius of the primary. Despite this fact, thanks to the higher
effective temperature of the primary, the relative luminosity of the primary component is about 4 times larger.
The scatter of the individual measurements is quite high, but the final fit is satisfactory.

\begin{table*}[b]
\small \caption{The light-curve parameters for PV~Cyg, V822~Cyg and V1011~Cyg, respectively.}
 \label{TableBig} \centering \scalebox{0.85}{
\begin{tabular}{c c c c c c c }
\hline
 Parameter         &          PV~Cyg           &       V822~Cyg           &       V1011~Cyg        \\
 \hline
 $JD_0$ $[$HJD$]$  & $2452595.252 \pm 0.002$   & $2452595.910 \pm 0.002$  & $2452598.790 \pm 0.001$  \\
 $P$ $[$day$]$     & $1.3148811 \pm 0.0000003$ &$1.2677742 \pm 0.0000004$ & $3.2393920 \pm 0.0000002$ \\
 $i$ [deg]         &   73.1 $\pm$ 1.2          &  89.8 $\pm$ 2.1          &  72.5 $\pm$ 0.9        \\
 $q_{ph} = M_2/M_1$ &  0.53 $\pm$ 0.05         &  0.35 $\pm$ 0.03         &  0.52 $\pm$ 0.09      \\
 $r_1 / r_2 $      &   0.71 $\pm$ 0.03         &  1.10 $\pm$ 0.01         &  0.54 $\pm$ 0.04     \\
 $T_1/T_2$         &   1.63 $\pm$ 0.06         &  1.99 $\pm$ 0.08         &  1.87 $\pm$ 0.10    \\
 $L_1 / L_2$       &   4.08 $\pm$ 0.21         &  9.20 $\pm$ 0.30         &  3.66 $\pm$ 0.08   \\
 $\Omega_1$        &   4.83 $\pm$ 0.11         &  3.14 $\pm$ 0.03         &  6.11 $\pm$ 0.24  \\
 $\Omega_2$        &   2.89 $\pm$ 0.05         &  2.44 $\pm$ 0.02         &  2.82 $\pm$ 0.19 \\ \hline
 \hline
\end{tabular}}
\end{table*}

\subsection{V822~Cyg}

The eclipsing binary V822~Cyg (= AN~216.1935, $R.A.=19\mathrm{^h}$~54$\mathrm{^m}$~08$\mathrm{^s}$,
$Decl.=+36\mathrm{^\circ}$~21$^\prime$~00\arcs, J2000.0, $B_{max} = 12.9$~mag) was discovered by
\cite{V822Cyg1935}. The light curve was observed and published by \cite{V822Cyg1959}, but any analysis has been
performed, so its parameters are still questionable. \cite{Svechnikov1990} published the first approximate
estimation of the spectral types, resulting in (A3)+[G2IV]. But this result is again not very reliable.

\begin{figure}
 \centering
 \includegraphics[width=13.8cm]{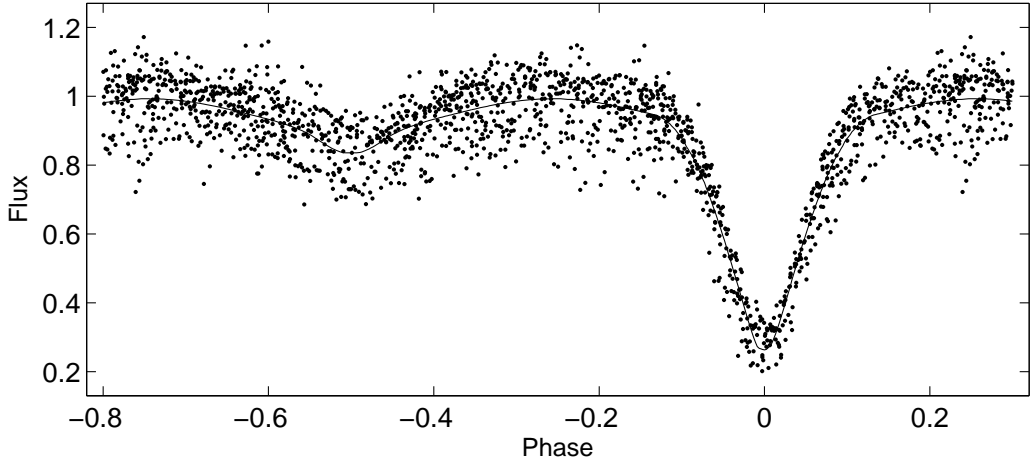}
 \caption{The $V$ light curve of V822~Cyg.}
 \label{FigV822Cyg}
\end{figure}

From these spectral types one could derive the temperatures $T_1 = 8600$~K and $T_2 = 5900$~K (according to
\citealt{Hec1988}). Exactly the same method as in the previous case was used. Altogether there are 1680
observations of V822~Cyg. The resulting parameters are in Table \ref{TableBig} and the plot is in
Fig.\ref{FigV822Cyg}.

Although the scatter of the measurements is rather high, thanks to the relatively large number of observations
(1855 in total), the final fit is satisfactory and no other effects are observable in this light curve.

\subsection{V1011~Cyg}

The last eclipsing binary selected for a detailed analysis is V1011~Cyg (= GSC~02677-01203,
$R.A.=19\mathrm{^h}$~55$\mathrm{^m}$~15$\mathrm{^s}$, $Decl.=+34\mathrm{^\circ}$~12$^\prime$~30\arcs, J2000.0,
$B_{max} = 12.2$~mag). It was discovered to be a variable by \cite{V1011Cyg1961}. The spectral type A0 was
presented in \cite{Brancewicz1980}, while more detailed classification as A0+[G3IV] was presented in
\cite{Svechnikov1990}.

\begin{figure}[h]
 \centering
 \includegraphics[width=14cm]{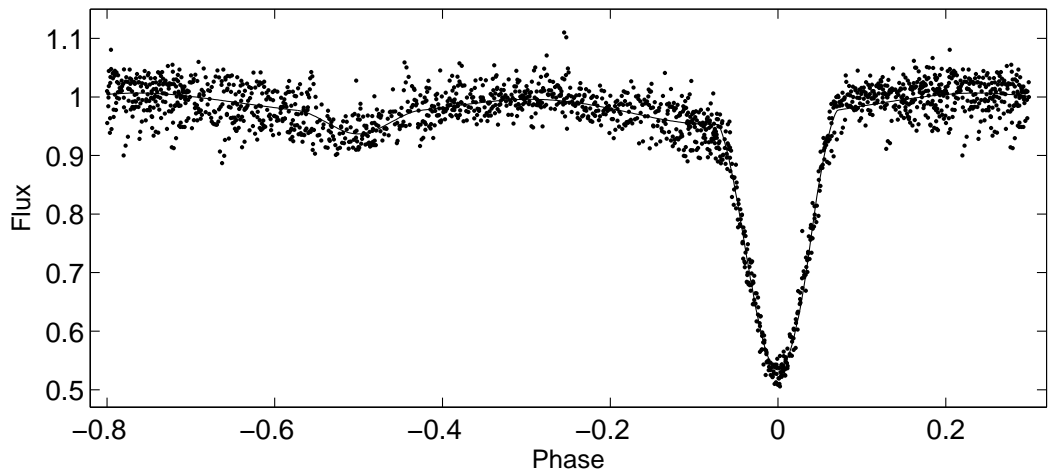}
 \caption{The $V$ light curve of V1011~Cyg.}
 \label{FigV1011Cyg}
\end{figure}

\begin{figure}
 \centering
 \includegraphics[width=12cm]{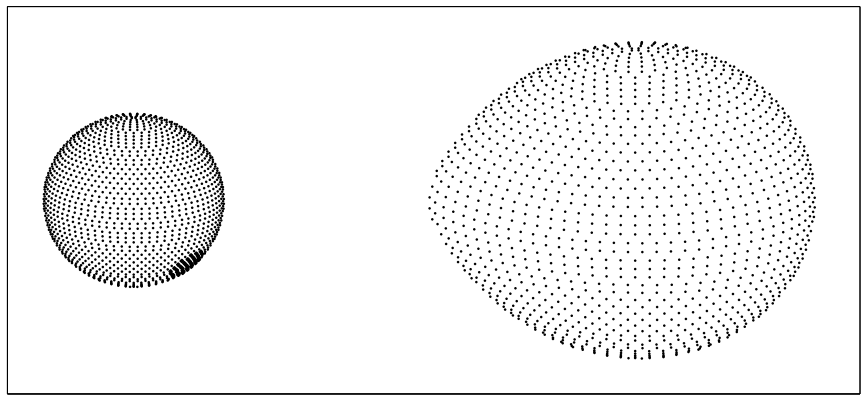}
 \caption{The 3-D plot of V1011 Cyg at the phase 0.25, primary is the left one.}
 \label{Fig3D}
\end{figure}

According to this latter paper, the proposed temperature of the primary is circa 9400~K, and about 5800~K of the
secondary. Altogether there are 1704 observations. As one can see, the curve is not symmetric and the
brightnesses near the phase -0.1 and near the phase 0.1 are not equal. This could be explained e.g. by the
presence of a spot. If one assume the spot to be located on the primary component, the parameters of such a spot
are the following:
\begin{center}
 \begin{tabular}{cc} \hline
 Parameter     & Value \\
 \hline
 Latitude  [rad] &  2.3 $\pm$ 0.9 \\
 Longitude [rad] & 5.61 $\pm$ 0.44 \\
 Radius [rad]    & 0.25 $\pm$ 0.04  \\
 $T_{spot}/T_{surface}$ & 1.7 $\pm$ 0.3 \\ \hline
 \end{tabular}
\end{center}
The final parameters of the light-curve fit are presented in Table \ref{TableBig} and the fit itself is plotted
in Fig.\ref{FigV1011Cyg}. The 3-D plot of the system is shown in Fig.\ref{Fig3D}.

However, the presence of spot on the primary is not able to describe the light curve in detail and at least two
spots are needed. One of them should be brighter and the other one dimmer than the surface of the star. There is
a slight difference between the shapes of the light curves obtained at different epochs. The first observations
from 2002 show larger difference in brightness near the phase -0.1 and 0.1, while the observations from the
following years show almost symmetric curve. This could indicate the time evolution of the spots. Only further
analysis would confirm the presence of the spots and their possible evolution.

\section{Discussion and conclusions}

The light-curve analysis of three selected systems in Cygnus constellation has been done. Using the light curves
observed by the INTEGRAL satellite, one can estimate the basic physical parameters of these systems. Despite
this fact, the parameters are still only the preliminary ones, affected by relatively large errors and many of
the relevant parameters were fixed at their suggested values. The detailed analysis is still needed, especially
in different filters. Together with a prospective radial-velocity study, the final picture of the systems could
be done. Particularly, the system V1011~Cyg seems to be the most interesting one due to its asymmetric light
curve.

\section{Acknowledgments}
Based on data from the OMC Archive at LAEFF, pre-processed by ISDC. This investigation was supported by the
Grant Agency of the Czech Republic, grants No. 205/06/0217 and No. 205/06/0304. We also acknowledge the support
from the Research Program MSMT 0021620860 of the Ministry of Education. This research has made use of the SIMBAD
database, operated at CDS, Strasbourg, France, and of NASA's Astrophysics Data System Bibliographic Services.

\end{document}